# Attention-Based Applications in Extended Reality to Support Autistic Users: A Systematic Review

**Katherine Wang[1,2], Simon Julier[3], and Youngjun Cho[1,2,3]**
[1]University College London Interaction Centre (UCLIC), University College London (UCL), London WC1E 6BT, United Kingdom
[2]Global Disability Innovation Hub (GDIH), University College London (UCL), London WC1E 6BT, United Kingdom
[3]Department of Computer Science, University College London (UCL), London WC1E 6BT, United Kingdom

**ABSTRACT**
With the rising prevalence of autism diagnoses, it is essential for research to understand how to leverage technology to support the diverse nature of autistic traits. While traditional interventions focused on technology for medical cure and rehabilitation, recent research aims to understand how technology can accommodate each unique situation in an efficient and engaging way. Extended reality (XR) technology has been shown to be effective in improving attention in autistic users given that it is more engaging and motivating than other traditional mediums. Here, we conducted a systematic review of 59 research articles that explored the role of attention in XR interventions for autistic users. We systematically analyzed demographics, study design and findings, including autism screening and attention measurement methods. Furthermore, given methodological inconsistencies in the literature, we systematically synthesize methods and protocols including screening tools, physiological and behavioral cues of autism and XR tasks. While there is substantial evidence for the effectiveness of using XR in attention-based interventions for autism to support autistic traits, we have identified three principal research gaps that provide promising research directions to examine how autistic populations interact with XR. First, our findings highlight the disproportionate geographic locations of autism studies and underrepresentation of autistic adults, evidence of gender disparity, and presence of individuals diagnosed with co-occurring conditions across studies. Second, many studies used an assortment of standardized and novel tasks and self-report assessments with limited tested reliability. Lastly, the research lacks evidence of performance maintenance and transferability. Based on these challenges, this paper discusses inclusive future research directions considering greater diversification of participant recruitment, robust objective evaluations using physiological measurements (e.g., eye-tracking), and follow-up maintenance sessions that promote transferrable skills. Pursuing these opportunities would lead to more effective therapy solutions, improved accessible interfaces, and engaging interactions.

## I. INTRODUCTION

Autism is a lifelong neurodevelopmental condition clinically defined by difficulties with social communication and interaction and by the presence of restrictive and repetitive behaviors and interests [1]. In reality, the autistic population is far more varied than may be gleaned from the listed criteria in the diagnostic manuals or the other widely used terminology (e.g., high-functioning, low-functioning). Complex differences in sensory sensitivities, the nature of repetitive behaviors, and the various types of all-encompassing interests all contribute to an autistic individual's ability to communicate. It is, therefore, more useful to consider autism as a constellation rather than a linear spectrum, as the latter over-simplifies this variability [2].

Attention is a complex cognitive process which involves and influences perception, memory, and decision-making to select aspects of information with which to interact [3]. Attention-related disorders, such as Autism Spectrum Disorder (ASD) have the potential to benefit from virtual environments. ASD is a pervasive neuropsychiatric diagnosis in which individuals experience hypo- and hyper-sensitivity towards physical stimuli due, in part, to over-selective attention [4]. Despite being easily affected by stimuli, individuals with ASD process visual cues more effectively than other types of sensory stimuli [5].

The umbrella term, extended reality (XR), encompasses a spectrum of augmented reality (AR) and virtual reality (VR) technologies that fall on the virtuality continuum [6]. It involves augmenting real-world content with virtual content and compelling digital cues using a camera and display, desktop monitor, head-mounted display (HMD), projection screen, or mobile device [7]. AR superimposes virtual information onto real-world content using a screen and camera [8] or a head-mounted display (HMD) [3], while VR allows users to interact with virtual environments [9].

XR allows researchers to control testing in a wide variety of scenarios and to produce more generalizable results due to increased ecological validity in comparison to traditional laboratory and clinical interventions.

Previous evidence indicates that performance improvements in a simulated environment transfer to the real-world for neurotypical individuals [10]. Similar results were also found for neurodiverse participants. For instance, improved attention performance achieved from a virtual classroom approach transferred to real-world classroom settings for participants with attention-deficit hyperactivity disorder (ADHD) [11]. Regardless of the variability between XR technology and intervention methods, cognitive and physical skills acquired during XR-based training transfer just as successfully to the real-world as those obtained from traditional training [12]. Furthermore, Kaplan *et al.* [12] argue that XR training provides benefits beyond traditional methods such as the ability to mitigate risk from dangerous situations.

Despite being hyper- or hypo-sensitive to stimuli, autistic individuals process visual cues more effectively than they do other types of sensory stimuli [5]. It is for this reason that XR can be particularly beneficial in supporting autistic users. AR, for instance, has been used to engage behavior by projecting information onto complex tasks to help users direct attention more efficiently [13]. Gaze-contingent interfaces have also been shown to facilitate interaction in joint attention tasks [14]; though, many of these effects are typically limited to reactions with the performance of neurotypical (NT) individuals. Due to the heterogeneity of autism, there is no particular technology or uniform intervention that can work for all users on the spectrum.

In order to inform researchers in the XR and autistic communities on promising directions for future exploration, we first provide a comprehensive overview of the methodologies used in a selection of studies. Next, we identify key methodological challenges of attention-based research for autism using XR and determine effective measurement tools and metrics. Lastly, this review proposes recommendations on more inclusive ways to conduct studies with autistic participants.

## II. BACKGROUND

### A. UNDERREPRESENTATION IN AUTISM RESEARCH

The global prevalence of diagnosed autism ranges between 0.3% to 1.2% of the population, with Western countries demonstrating a prevalence of 1% [15]. The majority of male participants in autism studies reflects the global male-to-female ratio in the identified autistic population (e.g., 1.33 to 16.0 in Northern Europe; 2.2 to 5.8 in North America), of which diagnosed males variably, but consistently outweigh diagnosed females [15]. However, recent findings suggest that autism presents differently in females and that current diagnostic criteria are largely male-biased [16] resulting in underdiagnosis of females and those who identify as other genders.

Autism research also focuses disproportionately on children due, in part, to the prioritization of understanding causal factors [17]. Furthermore, diagnostic criteria were far stricter during the advent of autism research in the 1960s and 1970s, and autistic adults who were not diagnosed or misdiagnosed at the time may have continued to manage thus far.

Lastly, there remains a controversial divide amongst communities, clinicians, and researchers between focusing on rehabilitation and developing medical cures, or on promoting acceptance and establishing rights [17].

### B. AUTISM AND ATTENTION

Table 1 describes the distinct attentional processes autistic people experience in comparison to neurotypical (NT) people (i.e., undiagnosed with ASD or do not display evidence of autistic traits). One such attentional process involves shifting focus from one aspect to the next or broadening and narrowing the area of focus (i.e., selective attention) [18]. For instance, in comparison to NT children, autistic children shifted attention more between non-social objects than between people [19].

Another attentional process relevant to autism includes the ability to focus on an activity or stimuli over a long period of time (i.e., sustained attention). Autism is most associated with differences in social attention (i.e., how focus is directed during a social situation). This includes joint attention (JA) (i.e., socially coordinated visual attention [18]), which plays a critical role in the development of social-cognitive processes involved in the inference of intentions or emotions of another person, as well as influencing how stimuli are encoded [20], [21].

Visual attention has been the most widely researched form of attention investigated in autistic populations. However, selected methodologies warrant more clarification as numerous conflicting findings are resulting from the use of methodologies that are not ideally suited to answer research questions or whose implications are not well understood [18].

The main paradigms used in visual attention research include eye-gaze cueing, spatial attention, eye-movement recording in scene-viewing tasks, change detection, and computational modelling [18]. Evidence also demonstrates different brain activity during attentional processing of sensory inputs (e.g., visual, auditory, tactile) [22]. Findings from other types of attention, specifically aural, generally correspond to those of visual attention [23]. These findings suggest that differences in attentional bias observed in autistic people during social and non-social situations may result from differences in brain activity, rather than neurological deficits.

**TABLE 1. Attention types**

| Attention Type | Description |
|---|---|
| Selective | • Increased perceptual capacity in autistic people compared to NT people can explain altered attention [23]<br>• Autistic individuals demonstrate strengths in non-search processes (e.g., discrimination) and reduced abilities in the generalization that is dependent on levels of perceptual load (i.e., the extent of distractor processing is dependent on the amount of perceptual load) [24]<br>• In terms of visual search, autistic adults demonstrated enhanced perceptual capacity during selective attention tasks when presented with greater perceptual load in comparison to non-autistic participants [25] |
| Sustained | • Autistic children perform similarly to NT peers on sustained attention tasks (e.g., Continuous Performance Test [26]) [27]<br>• Activity in the prefrontal cortex (associated with motivation and reward) differs between autistic and NT groups, implying that autistic people mediate motivation and respond to reward differently [28].<br>• Impairments in sustained attention may result from a misperception of disinterest or lack of motivation |
| Social | • During eye-tracking studies, autistic participants demonstrated differences in gaze patterns towards social regions (e.g., mouth, eyes) while viewing social scenes compared with NT groups [29]<br>• Autistic individuals demonstrated differences in initiating and responding to joint attention, which was linked to lower information processing in comparison to non-autistic peers [21]. The development of these is closely linked to the ability to learn with and from other people [30], suggesting that differences lead to learning problems in affected individuals [20]<br>• In autistic children, evidence of disrupted development of attentional bias towards social content resulted in missed opportunities to learn about social behavior and language [17] |

### C. RESEARCH GAPS

Previous systematic literature reviews analyzing research involving XR applications for autistic people [31]–[36] have all concluded that XR provided a safe and comfortable environment for autistic users to learn communication and social skills. However, half of these papers were focused on reviewing interventions designed for VR technology [34]–[36]. As a result, these reviews included limited amounts of literature, ranging from 6 articles [35] to 31 articles [33]. Several reviews also reached the common conclusion that there was a lack in methodological validity across studies which could be improved with more objective evaluation methods [31], [33], [36]. While existing review articles briefly discussed different types of evaluation parameters, none elaborated on the effectiveness of these parameters and the systematic synthesis of effective physiological and behavioral attention cues in autistic populations.

This systematic review aims to overcome this shortcoming by providing a comprehensive analysis of methodological aspects (i.e., study design, evaluations, limitations) and systematically assessing the effectiveness of physiological and behavioral metrics on attention components. This review is motivated, in particular, by supporting future XR research and development that identifies autistic characteristics as differences, rather than impairments or deficiencies, and works towards developing technology to support these differences across a variety of individuals.

## III. METHOD

Standard systematic review methods adopted from the Preferred Reporting Items for Systematic Reviews and Meta-Analyses (PRISMA) protocol [37] was used to conduct and report the current review.

### A. RESEARCH QUESTIONS

**TABLE 2. Research questions addressed in this systematic review.**

| Theme | ID | Question |
|---|---|---|
| Demographics | RQ1 | What countries did research originate from? |
| | RQ2 | What were the demographics of participants? |
| Study Design | RQ3 | What methods were used to screen participants during the selection process? |
| | RQ4 | Which XR technologies were used in the primary literature? |
| | RQ5 | Which attention types were targeted during primary literature? |
| | RQ6 | How was attention measured during the primary literature? |
| | RQ7 | What types of attention tasks were used during the primary literature? |
| Study Outcomes and Findings | RQ8 | How were XR-based attention tasks used most effectively and what were the outcomes? |
| | RQ9 | What were the attention outcomes of using affective measurements? |
| | RQ10 | What limitations and future research agendas were addressed in the primary literature? |

To better understand how XR studies can accommodate the differences in attention experienced by autistic individuals, this systematic review was guided by three themes (demographics of articles and participants, study design, and outcomes and findings) that include ten research questions in total. The first theme provides an understanding of how representative the participants are of the target population, while the second theme contributes data on methodological practices used in studies and the third theme addresses the effects of these methodologies on autistic participants. Altogether, these questions aim to demonstrate how much different research methods can be generalized and to shed light on alternative methods to improve generalization to different populations. Table 2 presents our detailed research questions under each theme.

### B. SEARCH STRATEGY.

A comprehensive literature search was conducted in June 2020 on six electronic databases, including ACM Digital Library, IEEE Xplore, ProQuest, ScienceDirect, Scopus, and SpringerLink. The search string displayed in Table 3 was developed based on the subjective judgement of the researchers of categorical terms that occurred frequently in preliminary literature searches. The terms included in the three identified categories (i.e., XR technology, autistic populations, or attention) were alternative names or related concepts. Search strings (Table 3) were tailored for each database and were used to search title, abstract, and author-specific keywords of articles. As XR is a relatively recent field, there was no limit to publication period to allow for more studies to be identified.

### C. SCREENING AND ELIGIBILITY

Articles were screened for duplicates using Mendeley software and verified for actual inclusion of the search terms within the full text. After duplicates were removed, titles and abstracts of the pooled articles were scanned, and full-texts were scanned in the following step.

**TABLE 3. Summary of search string strategy used in this systematic review.**

| Category | Search String |
|---|---|
| XR | ("virtual reality" OR "virtual environment" OR "VR" OR "augmented reality" OR "AR" OR "diminished reality" OR "DR" OR "mixed reality" OR "MR" OR "extended reality" OR "XR" OR "immersive technology" OR "three dimensional" OR "3D*" OR "3-D*") |
| | AND |
| Autism | (autis* OR ASD OR ASC OR Asperger OR "pervasive developmental disorder" OR PDD-NOS OR neurodevelopment* OR neurodiver* OR NDD) |
| | AND |
| Attention | (attention OR distract* OR concentrat* OR focus) |

For inclusion into the systematic review, titles and abstracts were scanned for articles focusing on autism and attention-related tasks (Table 4). Empirical studies were included where the interaction between humans and XR was part of an intervention that targeted autistic users.

Papers involving studies that did not include attention-based tasks using XR technology for autistic populations, or involved non-human participants, were excluded. Remaining papers were read in full and those that did not report on the design or effects of XR for autistic participants were also excluded.

**TABLE 4. Summary of inclusion, exclusion, and quality criteria of papers.**

| Criteria | Description |
|---|---|
| Inclusion | • Targets autistic users<br>• Reports outcomes related to attention, XR, and autism<br>• XR purpose or effects are, or can be isolated<br>• Articles involving empirical studies, evaluations, interventions, participatory design<br>• Full peer-reviewed text available in English |
| Exclusion | • Does not focus on attention-related aspects of autism<br>• Systematic reviews, scoping reviews, and meta-analyses<br>• Does not involve human users |

### D. DATA EXTRACTION

A data extraction form was developed to capture items related to study characteristics such as primary author, publication year, sample size, male-to-female ratio, age, diagnosis, XR hardware, characteristics of intervention, outcomes, limitations, and future research agendas.

An analysis of the selected literature was performed using data analysis software. Papers were imported and nodes were inductively created to analyse data from the methods, results, discussion, and conclusion sections of each paper according to our research questions. Similar to Stowell *et al.* [38] we iteratively coded and compared the literature using attributes (e.g., descriptions of participants, sample size of autistic participants, presence of NT participants, type of XR hardware used, intervention tasks, and attention evaluation measures, reported limitations) to determine emerging themes.

## IV. RESULTS

This section provides a concise overview of attention-based research for autism using XR. A total of 1531 articles were initially identified from the database search (Fig. 1). After removing duplicates, 1214 titles and abstracts were screened for XR-related keywords. A total of 1073 titles and abstracts were further screened against the exclusion criteria. The full texts of 173 articles were reviewed, of which 59 articles met the criteria for inclusion in the review and nine were used for a meta-analysis.

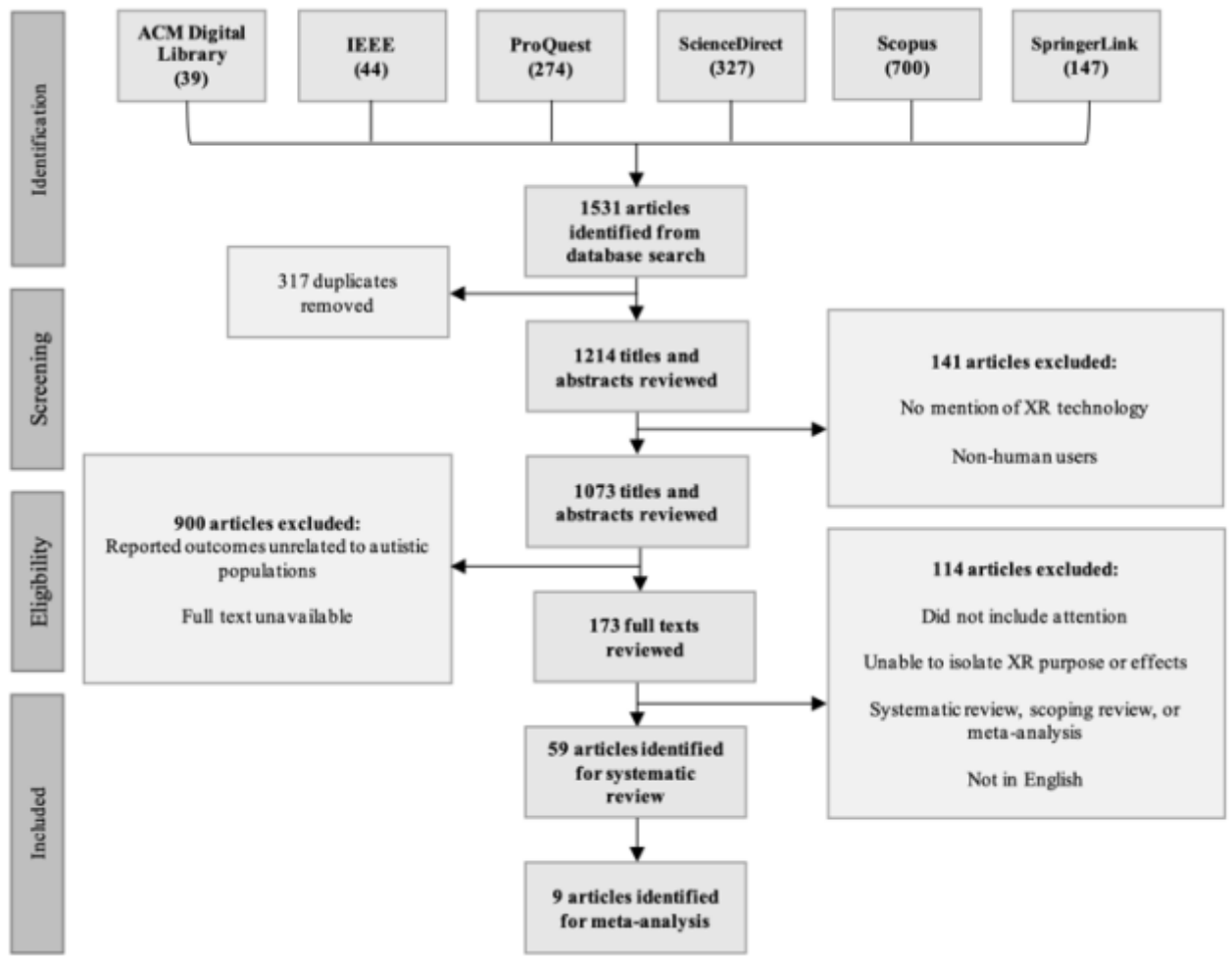

**FIGURE 1 Flowchart depicting the search and screening process of peer-reviewed articles from electronic databases**

**TABLE 5. Geographic locations of author affiliations.**

| Country | Lead Organization | Collaborating Organizations |
|---|---|---|
| United States | 27 | 29 |
| India | 6 | 5 |
| China | 4 | 2 |
| United Kingdom | 4 | 4 |
| Taiwan | 3 | 3 |
| Indonesia | 2 | 4 |
| Japan | 2 | 3 |
| Portugal | 2 | 2 |
| Spain | 2 | 2 |
| Canada | 1 | 1 |
| Ecuador | 1 | 1 |
| France | 1 | 1 |
| Hong Kong | 1 | 2 |
| Italy | 1 | 1 |
| Qatar | 1 | 1 |
| Turkey | 1 | 1 |
| Mexico | - | 1 |
| Scotland | - | 1 |
| Wales | - | 2 |
| Australia | - | 1 |
| Belgium | - | 1 |
| South Korea | - | 1 |
| Sweden | - | 1 |
| Switzerland | - | 1 |

### A. RQ1: RESEARCH AFFILIATIONS?

The 59 papers reviewed work led by first authors affiliated with academic or professional institutions across 16 different countries (Table 5). There were also 262 collaborating authors from 71 academic or professional institutions across 24 different countries. The United States had substantially more affiliations than all other countries, followed by India, the United Kingdom, China, Taiwan, and Indonesia. We did not identify any papers affiliated with African countries.

### B. RQ2: WHAT WERE THE DEMOGRAPHICS OF PARTICIPANTS?

Out of 59 papers, 54 papers involved autistic participants, totaling 736 participants. The following section covers the different characteristics of these participants. Characteristic distributions were only calculated from papers that provided a sample size of autistic participants.

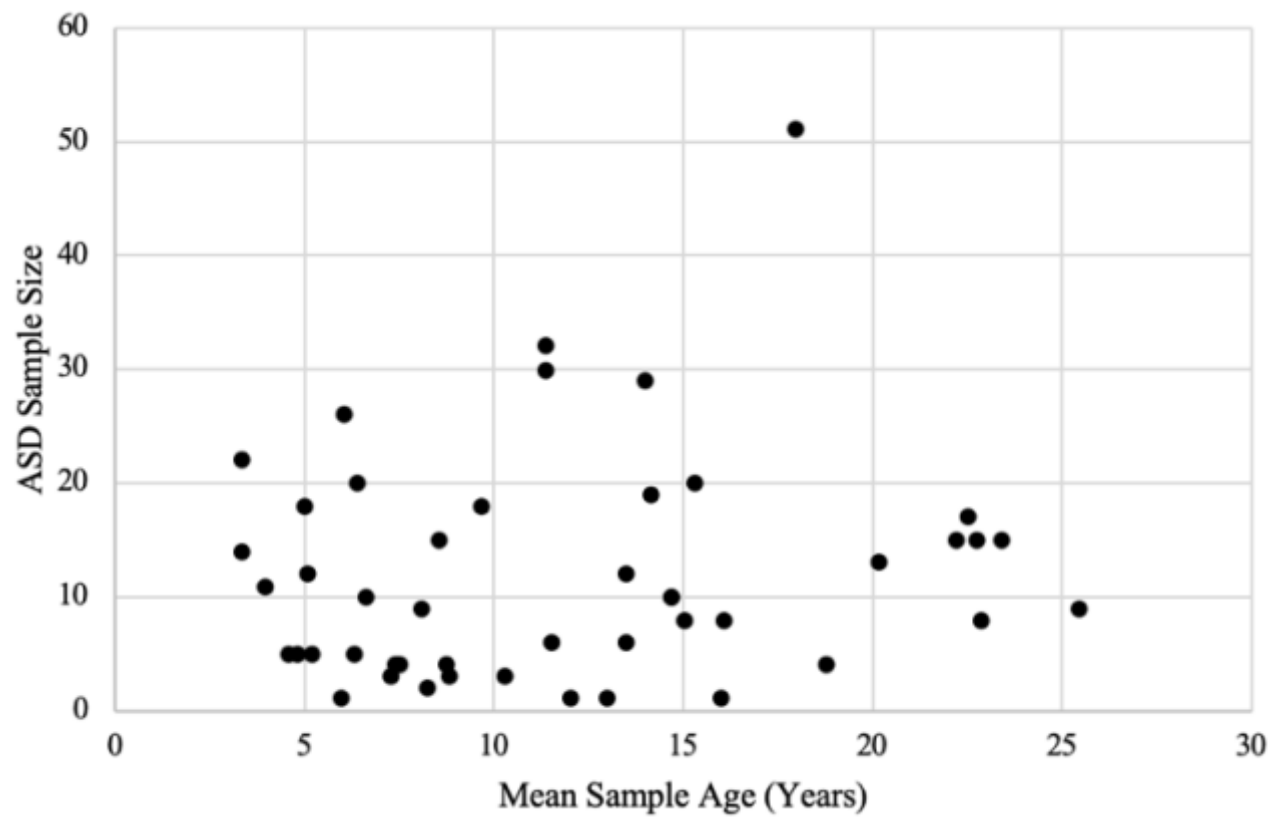

**FIGURE 2. Average age and sample size of autistic participants**

43 papers reported the average age for their sample of ASD participants (Fig. 2). The 559 participants ranged from 3.33-25.44 years of age (μ=11.78, σ=6.17), and the average sample size ranged from 1 to 51 participants with ASD (μ=11.89, σ=9.97).

Among the studies that did not disclose the average age for included ASD participants, six papers highlighted studies that involved participants with an age-range under 18 years [39]–[44] while one paper involved adult participants (i.e., over 18 years) [8] and four papers failed to report the average age of autistic participants [45]–[48]. Five papers proposed interventions intended for autistic users without involving autistic participants in studies. In regards to these papers, three presented interventions intended for autistic children [49]–[51] while two others did not specify the age range of target users in their studies [52], [53]. None of the interventions were designed for autistic adults.

Gender distributions were calculated from autistic participants. 35 papers reported participants as either male or female, while 18 papers did not report gender. Of the 440 participants for whom gender was reported, 87.7% of them were male. This echoes previous findings showing that the majority of participants in autism studies are male. None of the studies in the current review acknowledged other gender identities (e.g., non-binary, gender-fluid).

### C. RQ3: WHAT METHODS WERE USED TO SCREEN PARTICIPANTS DURING THE SELECTION PROCESS?

Out of the 54 papers with ASD participants, 14 papers had no indication of autism type while 33 papers required participants to have received some form of autism assessment. As displayed in Table 6, the following diagnostic tools were used across studies: Autism Diagnostic Observation Schedule [54] (ADOS), Social Responsiveness Scale [55] (SRS), Social Communication Questionnaire [56] (SCQ), Childhood Autism Rating Scale [57] (CARS), Autism Diagnostic Interview-Revised [58] (ADI-R), Autism Spectrum Screening Questionnaire [59] (ASSQ), Gilliam Autism Rating Scale [60] (GARS), Autism Spectrum Inventory [61] (ASI), and Autism Quotient-Child (AQ-Child). 21 papers also reported using some form of intelligence (IQ scores e.g., full-scale IQ, verbal IQ, performance IQ). It is also worth noting that all five papers presenting preliminary designs and studies without ASD participants did not indicate diagnostic traits for their target audience [40], [49]–[51], [53].

**TABLE 6. Evaluation tools used to screen participants for studies.**

| Diagnostic Tool | Papers |
|---|---|
| **Autism Diagnostic Observation Schedule** [54] (ADOS) *(12)* | [62]–[73] |
| **Social Responsiveness Scale** [55] (SRS) *(11)* | [55], [63], [64], [66], [67], [70], [74]–[78] |
| **Social Communication Questionnaire** [56] (SCQ) *(9)* | [56], [66], [67], [74], [75], [77]–[80] |
| **Childhood Autism Rating Scale** [57] (CARS) *(4)* | [73], [81]–[83] |
| **Autism Diagnostic Interview-Revised** [58] (ADI-R) *(4)* | [62], [66], [71], [82] |
| **Autism Spectrum Screening Questionnaire** [59] (ASSQ) *(3)* | [59], [75], [79] |
| **Gilliam Autism Rating Scale** [60] (GARS) *(2)* | [45], [84] |
| **Autism Spectrum Inventory** [61] (ASI) *(1)* | [85] |
| **Autism Quotient-Child** (AQ-Child) *(1)* | [76] |
| **Intelligence Quotient (IQ) Scores** (e.g., full-scale IQ, verbal IQ, performance IQ) *(21)* | [43], [48], [62]–[64], [67]–[69], [71]–[73], [75], [76], [79], [81]–[83], [86]–[89] |
| **Not specified** *(14)* | [8], [39], [40], [42], [44], [47], [90]–[97] |

### D. RQ4: WHICH XR TECHNOLOGIES WERE USED IN THE PRIMARY LITERATURE?

46 out of 59 papers involved VR technology, with applications deployed on computer monitors ($n$ = 18) or HMDs ($n$ = 17) (e.g., Oculus Rift [42], [49], [71], [72], [91], [92], Google Cardboard [93], [98], HTC Vive [53]) being the most popular. Other VR modalities include screen projections (e.g., [89], [94], [99]), CAVEs [41] and multi-touch displays ([39], [80]).

AR technology was used in 13 papers. HMDs (e.g., Google glass [77], [97], [100]) were also popular among AR studies, along with mobile devices (i.e., smartphone or tablets [8], [85], [88]). Additional AR modalities included laptop [48] or tablet PC [87], diminished reality interface [52], or large-scale floor projection [43].

### E. RQ5: WHICH ATTENTION TYPES WERE TARGETED DURING PRIMARY LITERATURE?

As displayed in Table 7, nearly half of the papers included in this review focused on social attention. Four papers [49], [70], [101], [102] focused on selective and sustained attention. Two articles [51], [89] focused on social, selective, and sustained attention.

Of the included literature, significantly more focused on social attention and involved two or more co-occurring tasks (33.3%), compared to those that investigated only selective attention (18.8%) or only sustained attention (17.6%) studies.

**TABLE 7. Context of attention targeted in studies.**

| | |
|---|---|
| **Social** *(30)* [41], [44], [47], [48], [49], [52], [61], [62], [64], [65] [68]–[71], [73]–[78], [79], [81], [85], [86], [88], [92], [93], [95], [97], [98], [99] | **Selective** *(15)*<br>[8], [40], [43], [46], [49], [50], [67], [82], [88], [90], [91], [94], [100]–[102]<br>**Sustained** *(16)*<br>[39], [40], [44], [46], [49], [63], [66], [67], [72], [88], [89], [91], [96], [100], [101], [103] |

### F. RQ6: HOW WAS ATTENTION MEASURED IN THE PRIMARY LITERATURE?

Overall, 52 of the 59 papers mentioned using attention measurement strategies with autistic participants. The two categories of studies included quantitative and qualitative. Physiological and behavioral signatures were used in 22 papers making them the most frequently used measures across the featured literature (Table 7), We found eye tracking was used exclusively in 72.7% of these studies. Motion-tracking sensors were also widely used (63.6%) to measure body position ($n$ = 7), hand gestures ($n$ = 5), and head position ($n$ = 3). Additionally, several papers also measured brain activity using functional magnetic resonance imaging (fMRI) ($n$ = 2) or electroencephalogram (EEG) ($n$ = 4), with two also employing eye-tracking in tandem ([69], [71]). Three papers (13.6%) also applied other physiological measurements (e.g., electrocardiogram, electrodermal activity, electromyogram, galvanic skin response, pulse plethysmogram, skin temperature, respiration).

Scored assessments ($n$ = 13) include standardized assessments (e.g., [85]: ASI, NEPSY-II [68]: NEPSY-II, [100]: ABC-H), or self-developed rating scales (e.g., 'JAAT' [72]). Performance-based measurements ($n$ = 11) encompass reaction time, task completion, and frequency. Fewer papers involved studies relying on qualitative measurements, such as observations and interviews (17.3%).

### G. RQ7: WHAT TYPES OF ATTENTION TASKS WERE USED DURING THE PRIMARY LITERATURE?

The types of attention-based tasks implemented in each paper were dependent on aims of the study. These included improving attention or training a skill (e.g., pronunciation, emotion recognition) or assessing attention differences between autistic and NT participants or amongst different types of technology.

**TABLE 8. Types of physiological and behavioral signals measured in studies.**

| Physiological and behavioral cues | Papers |
|---|---|
| **Eye movement (mainly fixation)** (*n* = 18) | [45], [49], [50], [53], [63], [65]–[67], [70], [72], [76], [77], [79], [82], [90], [92], [94], [103] |
| **Body position** (*n* = 7) | [42], [43], [84], [89], [91], [95], [96] |
| **Hand gesture** (*n* = 5) | [47], [51], [86], [96], [99] |
| **EEG** (*n* = 4) | [53], [70]–[72] |
| **Head position** (*n* = 3) | [69], [75], [91] |
| **Brain activity** (*n* = 2) | [62], [64] |
| **Facial expressions** (*n* = 2) | [41], [101] |
| **Other affective signals** (*n* = 3) | [63], [70], [78] |

Tasks involving imitation, interaction with an avatar or agent, following cues, and contextual familiarization were primarily used to for improving attention or training skills. Interacting with objects and recognizing and identifying objects, emotions, or words were also used for these aims, apart from [65], [67], [74], [79] which compared differences between autistic and NT participants and [66], [74], [95] which compared attention between different modalities. Tasks requiring participants to initiate or maintain eye gaze or passively attend to a scene were used for different kinds of aims.

Table 8 shows the different types of attention-based tasks used in the literature. While most papers involved interaction with either a virtual agent or object, these interactions often occurred with a second gaze-based task, in which participants fixated eye-contact onto stimuli [72], [73], followed cues to interact with objects [8], [78], [96] or to maintain gaze [49], or interacted with an avatar to identify emotions [64]. Two studies involving four tasks (e.g., interact and maintain gaze with an avatar to follow cues to interact with objects) [50], [51] focused on social, selective, and sustained attention.

### H. RQ8: HOW DID XR-BASED TASKS TARGET ATTENTION AND WHAT WERE THE OUTCOMES?

The main purpose of XR-based tasks intended to improve attention or attention-related skills through intervention. Other tasks served to help researchers understand the differences in attention between autistic and NT participants. We identified eight domains of tasks (Table 9) that involved responding to social cues from an avatar or agent (e.g., Maintain/initiate gaze, imitation), interacting in realistic scenarios (e.g., Contextual familiarization), or playing games involving cognitive tasks (e.g., Recognition/identification). Table 10 further specifies tasks and their outcomes for each paper.

**TABLE 9. Task domains performed during studies.**

| | |
|---|---|
| **Maintain/initiate gaze** (*n* = 9) [39], [45], [49]–[51], [72], [73], [75], [82] | **Interact with objects** (*n* = 13) [8], [40], [43], [48], [50], [51], [65], [78], [80], [83], [89], [95], [96] |
| **Imitation** (*n* = 3) [80], [87], [104] | **Contextual familiarization** (*n* = 2) [89], [93] |
| **Interact with avatar or other** (*n* = 15) [42], [45], [50], [51], [53], [64], [66], [68], [72]–[74], [77], [79], [80], [97], [100] | **Other** (*n* = 8)<br>Passive (just attend to a scene) [44], [62], [90], [99]<br>avoid distraction: [101]<br>Driving: [70], [94]<br>Respond to audio: [91]<br>Choose preferred stimuli: [76]<br>Stroop: [69] |
| **Recognition/identification** (*n* = 7)<br>Objects [81], [102], [105]<br>Emotions [63], [64], [67]<br>Words [73] | |
| **Following cues** (*n* = 10) [8], [49]–[51], [71], [77], [78], [87], [96], [98] | |

#### RESPONDING TO AFFECTIVE CUES FROM AN AVATAR/AGENT

Seven papers explored tasks requiring participants to select target items indicated by different social cues from avatars/agents (e.g., shifting gaze, pointing, head-turning) ([45], [50], [51], [71], [72], [78], [79]). In one study, all participants successfully followed avatar head movements by shifting their gaze to identify target objects placed around a virtual bedroom [71]. In another study, participants initiated eye contact with an avatar and responded to gaze shifts towards an image [79]. Authors of this study found that, compared to participants who were neurotypical or diagnosed with ADHD, autistic participants were less likely to exhibit effects associated with initiating joint attention on information processing. Jyoti and Lahiri [78] responded to different social cues from an avatar by tapping on the object of interest. Researchers found pointing to an object as the easiest cue for autistic children to decipher, in comparison to shifting eye-gaze or head-turning, which were found to be equal in effectiveness for NT children. However, Amaral *et al.* [72]

**TABLE 10. Summaries of methodology and outcomes of literature featured in this systematic review.**

| Reference | Performance evaluation | Task involving XR | Attention measurement | Outcomes |
|---|---|---|---|---|
| **Strickland *et al.*** [81] **(1996)** | Determine tolerance for VR equipment | Watch moving car or follow images in VR HMD | Engagement – Verbal responses | Similar responses between virtual scenes; meaningful engagement with VR |
| **Mineo *et al.*** [39] **(2009)** | Examine engagement with different electronic screen media | (1) Play VR game or (2) watch another person play a VR game | Engagement – Observations of gaze duration and verbal responses | Other VR: Increased vocalization<br><br>Self-VR: Increased gaze duration |
| **Pitskel *et al.*** [62] **(2011)** | Investigate differences in processing of direct and averted gaze | View a realistic social scenario in VR of an approaching person with (1) direct gaze, or (2) averted gaze | Brain activity in regions associated with attention | ASD group: active to averted gaze<br><br>NT control group: active to direct gaze |
| **Cheng and Huang** [86] **(2012)** | Train joint attention skills | Respond to 10 randomly selected pictures using desktop VR and a data glove | Observations of joint attention skills (e.g., pointing, showing, sharing); assessment scale scores adapted from [106] | Improvements in joint attention skills |
| **Grynszpan *et al.*** [82] **(2012)** | Monitor visual attention during a conversational context | Look at virtual face in desktop VR while receiving visual feedback on gaze-direction<br><br>Baseline and Final conditions: No feedback | Quantity of gaze fixations and sum of gaze durations; Questionnaire scores | Higher scores in final condition than baseline / experimental<br>Greater fixations on surroundings vs. face<br>Significant differences in average duration of gaze fixations during experimental condition |
| **Bekele *et al.*** [63] **(2013)** | Evaluate efficacy of VR to train emotion identification | Watch and identify emotional facial expressions | Gaze pattern towards different facial ROIs (e.g., forehead, eyes, nose, mouth) - Gaze patterns on specific ROIs | More randomized distribution tendencies of gaze patterns than NT group; significant gaze differences for mouth and forehead ROIs |
| **Jarrold *et al.*** [75] **(2013)** | Assess joint attention during a public speaking task | Answer self-relevant questions while looking at virtual avatars | Quantity of gaze fixations and sum of gaze durations measured using head-position tracking | ASD with high social anxiety linked to lower orientation in high ADHD inattention groups or lower IQ groups; No significance differences with low social anxiety groups |
| **Lahiri *et al.*** [66] **(2013)** | Investigate differences in gaze-patterns between systems: (1) engagement-sensitive and (2) performance-sensitive | Interact with agent using desktop VR during and after them giving a presentation | Social task performance, gaze patterns on ROIs and eye-physiology | Improved social task performance |
| **Wang and Reid** [83] **(2013)** | Investigate efficacy of VR as a cognitive rehabilitation tool | Move objects of different characteristics (e.g., color, dimensions) to matching contexts | VR Test of Contextual Processing of Objects [107]<br>Flexible Item Selection Task [108]<br>Control: Attention Sustained Subtest [109] | Improved contextual processing and cognitive flexibility; attention assessment scores same for nearly all participants |
| **Bekele *et al.*** [67] **(2014)** | Test responses during facial affect recognition | Identify different emotions using desktop VR | Affect recognition accuracy, gaze patterns | More randomized distribution tendencies of gaze patterns than NT group; significant gaze differences for mouth and forehead ROIs |
| **Escobedo *et al.*** [102] **(2014)** | Explore how mobile AR can help selective and sustained attention | Interact with digital content superimposed over physical objects | Observations of reactions | 62% increase in selective attention and 45% increase in sustained attention while using application |
| **Wang *et al.*** [49] **(2014)** | Test VR system's ability to train eye-contact with NT participants | Watch videos of virtual objects and displayed prompts | Gaze duration on ROIs | Fading prompts are most effective at drawing attention of NT participants |
| **Chen *et al.*** [87] **(2016)** | Use desktop AR to attract attention and strengthen focus during social situations | Focus on nonverbal social cues | Interview on observed emotions | Increased attention towards social aspects of images |
| **Muneer *et al.*** [40] **(2015)** | Explore efficacy of VR games for cognitive, social, and motor skills | Play games that require interaction with virtual objects using hand, arm, and body movements | Fixation duration and ability to pay attention to the game while avoiding distractions | Posttest improvements in sustained and selective attention |
| **Yantaç *et al.***[52] **(2015)** | Propose a diminished reality interface | Participate in a co-design workshops to help develop interface | - | - |
| **Didehbani *et al.*** [68] **(2016)** | Investigate efficacy of desktop VR for social cognition training | Practice social skills in different scenarios | Assessment – NEPSY-II Auditory Attention and Response Set | Improvement in assessment scores |
| **Lorenzo *et al.*** [41] **(2016)** | Test ability of application to detect and train emotional responses | Practice social skills for different scenarios | Ability to pay attention to social scenarios | Improvements related to appropriate emotional behaviors and compliance with guidelines |
| **Mundy *et al.*** [79] **(2016)** | Examine effects of joint attention on information processing using VR | Initiate gaze with avatar | Fixation location and gaze duration | ASD less likely than participants with ADHD to exhibit joint attention effects |
| **Parsons and Carlew** [69] **(2016)** | Compare distractions in desktop VR to traditional and computerized modalities | Perform Stroop task using different modalities | Ability to perform Stroop task | Worse performance than NT in VR; similar performance in traditional and computerized modalities |
| **Wade *et al.*** [70] **(2016)** | Conduct primary assessment of gaze-contingent driving simulator in desktop VR | Perform driving simulations | Task duration and number of failures to follow traffic signals | Significant improvements in performance post-test compared to pre-test |
| **Winoto** [105] **(2016)** | Improve word recognition using VR | Focus on an object and listen to its name being pronounced | Observations of usability | - |
| **Amaral *et al.*** [71] **(2017)** | Investigate feasibility of VR brain computer interface (BCI) to train social cognition skills | Respond to agent's head turns by identifying object of interest | Accuracy of BCI to identify attention markers as target event; accuracy of participant in following joint attention cues | BCI correctly identified attention markers and participant was able to read joint attention cues correctly |
| **Bozgeyikli *et al.*** [89] **(2017)** | Investigate effectiveness of VR system to train transferrable vocational skills and test distracter effects | Learn and practice vocational skills in different scenarios | Level score, accuracy, types of distractors encountered, number of prompts required, completion time | Improvements in all vocational skills and no negative effects of distracters on performance |
| **Cox *et al.*** [94] **(2017)** | Investigate differences of autistic drivers and efficacy of VR driving simulator to improve performance | Perform driving simulations | Ability to follow traffic signals | Better baseline performance resulted in greater improvements in performance |
| **Naranjo *et al.*** [42] **(2017)** | Implement and test a VR-assisted teaching process compared to traditional methods | Attend taught class in VR with a physical robot toy | Observations of system's ability to capture and maintain attention | Better ability to maintain attention in VR compared to traditional methods |
| **Ip *et al.*** [95] **(2017)** | Examine if attentional functioning is similar in 2D vs. 3D VR tasks | Indicate when moving stimuli touches target by raising arm | Visuospatial attention (accuracy and RT of movements)<br>ANT-C | No significant difference between ASD and NT groups in 2D task. The ASD group showed similar accuracy but slower reaction time than NT group in 3D task |
| **Amat *et al.*** [50] **(2018)** | Present design of VR game to improve joint attention | Follow avatar's eye and head movements to solve a puzzle | Accuracy and completion time | Easy to understand and improved performance over time<br>*NT participants only |
| **Amaral *et al.*** [72] **(2018)** | Investigate feasibility of VR brain computer interface to train social cognition skills | Respond to avatar's gaze towards an object by looking at it | Detection of social cues from avatar (i.e., gaze direction, pointing) | No significant improvements |
| **Bozgeyikli *et al.*** [96] **(2018)** | Explore effects of instructional methods (Exp. 1), visual fidelity (Exp. 2), and clutter and motion (Exp. 3) in VR warehouse training | Follow cues to find, position and move boxes onto moving targets | Experiment 1: Amount of focus on preferred instruction method<br>Experiment 2: Ability to follow moving stimuli<br>Experiment 3: Locating stimuli among clutter | Experiment 1: Preference for animated instructions<br>Experiment 2: Visual fidelity had a significant impact on attention<br>Experiment 3: Clutter significantly affects selective attention |
| **Feng *et al.*** [76] **(2018)** | Examine uncanny valley effects using desktop VR | Indicate preference for displayed faces on a screen | Fixation duration on facial ROIs measured using eye-tracking | No significant difference in attentional preference between faces for ASD group (i.e., absence of uncanny valley effect) |
| **Keshav *et al.*** [97] **(2018)** | Assess teacher response of AR smartglasses intervention in attention and social educational learning | Engage in guided conversations with a teacher while wearing smartglasses | Observational reports of ability to maintain attention | High attention levels during conversations |
| **Lee *et al.*** [88] **(2018)** | Train responses to nonverbal social cues with concept mapping using tablet AR | Focus on social cues and answer questions about appropriate greeting behaviors | Ability to follow instructions to answer questions | Increased response accuracy |
| **Manju *et al.*** [90] **(2018)** | Propose a therapeutic framework to enhance attention using VR | Explore virtual environment with colorful lights and sounds, interact with peer | Attention grasping and social interactions | Improved posttest scores on attention grasping and social interaction |
| **Mei *et al.*** [45] **(2018)** | Investigate effects of customizable avatars in desktop VR to train joint attention | Follow gaze of virtual teacher and view surroundings | Number of times joint attention is established, RT, and gaze duration measured by eye-tracking | More time spent gazing at relevant ROIs and greater RT when interacting with customizable virtual humans |
| **Nazaruddin and Efendi** [46] **(2018)** | Investigate feasibility of pop-up AR books to improve object focus and recognition | Interact with the prototype | Observations and interviews regarding curiosity and mastery of message content | Increased interest in design and mastery of message content |
| **Porayska-Pomsta *et al.*** [80] **(2018)** | Examine efficacy of virtual environments in VR for learning and improving social skills | Rehearse joint attention with an agent in different activities | Observations of initiating and responding joint attention | Increased initiations of joint attention during intervention |
| **Saadatzi *et al.*** [73] **(2018)** | Examine effectiveness of a learning environment using desktop VR and a robot peer | Follow social cues from virtual teacher to learn to read with positive feedback | Ability to respond to social cues and learn words | Acquired, maintained, and generalized 100% of taught words |
| **Syahputra *et al.*** [47] **(2018)** | Train focus using AR image capture and marker system | Move an object through paths to collect coins using hands | Sustained attention | Positive evaluation on system's ability to train focus |
| **Takahashi *et al.*** [43] **(2018)** | Develop a large-scale AR system for learning through empathic design | Follow and stand on moving objects projected onto a floor | Observations of attention to projections | Weak, but promising findings of attention |
| **Tang *et al.*** [99] **(2018)** | Study usability and engagement levels of gesture-based games in desktop VR | Study 1: Perform drawing gestures<br>Study 2: Pair words with corresponding images | Observations of attention levels towards virtual objects | High sustained attention levels during game and preference for touch-based interfaces |
| **Vahabzadeh *et al.*** [77] **(2018a)** | Investigate efficacy of AR smartglasses on irritability, hyperactivity, and social withdrawal | Interact with a facilitator while receiving cues from AR system | Inattention and distractibility – ABC scores | Decreased inattention and distractibility at end of weeks 1 and 2 compared to control week |
| **Vahabzadeh *et al.*** [100] **(2018b)** | Access short-term effect of intervention with AR smartglasses on ADHD-related symptoms | Interact with caregiver using AR system | Inattention and distractibility – ABC-H scores | Decreased inattention levels 24 and 48 hours after intervention |
| **Yang *et al.*** [64] **(2018)** | Investigate efficacy of behavioral interventions through neural functioning using VR | Interact with clinician through avatar in contextual scenarios | Changes in blood oxygen levels in neural regions associated with visual attention | Decreased brain activity towards nonsocial stimuli; preference for nonsocial features over social features |
| **Banire *et al.*** [101] **(2019)** | Evaluate facial expressions during attention tasks using desktop VR | React to cues on virtual chalkboard while avoiding distractions | Accuracy of reactions and tracked facial expressions | Prominent expressions related to attention were found to include raised brow, opened mouth, and lip sucking |
| **Chen *et al.*** [103] **(2019)** | Evaluate ROIs on a virtual tutor and efficacy of desktop VR system to train pronunciation | Mimic tutor | Gaze duration on ROIs measured by eye-tracking | |
| **De Luca *et al.*** [84] **(2019)** | Test efficacy of VR in cognitive behavioral therapy training | Identify, find, count, describe, chase, or move objects in virtual environment | Selecting randomly appearing elements | Significant increase of attention processes using VR CBT training |
| **Johnston *et al.*** [91] **(2019)** | Examine auditory spatial attention and sound localization using VR HMD | Locate sound-emitting stimuli in a virtual environment | Localization time, positional data accuracy | Higher performance in binaural-based spatial audio groups; longer localization times when background noise is present |
| **Koirala *et al.*** [65] **(2019)** | Investigate effectiveness of desktop VR to assess sensory processing differences | Move a virtual ball along a path using a haptic arm | Distraction levels (gaze towards distractors), position of stimuli with haptic feedback | ASD group focused more on distractions than NT group, but no significant differences in performance |
| **Kumazaki *et al.*** [74] **(2019)** | Examine differences in responses to social bids from different types of technological agents | View an agent in desktop VR | Frequency of head-turns or eye-gaze toward virtual agent measured by observation | ASD group-oriented head more toward the simple robot and virtual avatar than other modalities |
| **Lorenzo *et al.*** [85] **(2019)** | Assess effectiveness of mobile AR intervention on social skills | Complete social activities (e.g., interact with avatar, follow prompts to identify objects) | Autistic Spectrum Inventory | ASD group showed slight improvements in flexibility and imitation skills |
| **Rahmadiva *et al.*** [51] **(2019)** | Test multipurpose VR HMD games that train focus | Sort objects, follow signage, and maintain eye contact with avatar | Task completion time and accuracy | Increased accuracy with sorting objects, decreased completion time in following signage, similar incorrect answers with avatar |
| **Ravindran *et al.*** [98] **(2019)** | Evaluate feasibility of joint attention training application using VR HMD | Learn joint attention behaviors from instructional cards | Observations of shifting eye-gaze, initiating and responding to interactions, pointing behavior | Increased eye contact and initiating eye-gaze |
| **Rosenfield *et al.*** [92] **(2019)** | Validate feasibility and acceptability of VR system | Purchase fish from an agent using HMD | Coded observations of interaction and verbal exchanges and interviews | Participants were interested and easily engaged by the interface |
| **Singh *et al.*** [48] **(2019)** | Evaluate application of desktop AR system to enhance learning | Complete a tangram puzzle with instructions | Completion time, accuracy, and understanding measured by a questionnaire | Better performance in in-person modes compared to with AR system; inability to complete puzzles without cues |
| **Suresh and George** [44] **(2019)** | Investigate efficacy of VR HMD to distract children from anxiety during dental visits | Watch virtual content while attending a real-world dental visit | Distraction levels – Anxiety and behavior measured by assessments (Venham's Picture Test, Frankel's Behavior Rating Scales) | Increased distraction from dental visits while using VR HMD |
| **Jyoti and Lahiri** [78] **(2020)** | Investigate implications of different joint attention cues using desktop VR | Follow joint attention cues to select target objects on a touchscreen | Response accuracy, RT, distribution of screen taps on ROIs, physiological signals (PPG, EDA) | Improved performance for all types of joint attention cues |
| **Kelly *et al.*** [53] **(2020)** | - | - | - | - |
| **Miller *et al.*** [93] **(2020)** | Train air travel skills using VR HMD | Perform virtual rehearsal of air travel situation | Pre- and post-test subjective questionnaire by parents; clinical observations | Increase in attention to VR intervention associated with increase in use of targeted vocabulary |
| **Wang *et al.*** [8] **(2020)** | Propose an attention-training interface | - | - | - |

found no difference in participant ability to identify target objects from avatar cues accurately (e.g., pointing, head-turning).

Two articles included studies examining responses to viewing social cues [62], [82]. Viewing a social scenario scene was used to distinguish attention patterns and between ASD and NT participants. When viewing a realistic scenario of an approaching male figure, [62] found activation in neural regions associated with social attention (i.e., right temporoparietal junction) to averted gaze in autistic participants, compared to activation to direct gaze in NT participants. In another study, researchers examined viewing patterns towards a realistic virtual face using a gaze-contingent display in which everything but the focal point was blurred [82]. Researchers noted highly heterogenous viewing patterns among autistic participants but found they generally focused on areas surrounding the face more than NT participants. Also, while the gaze-contingent lens resulted in more gaze stability, only one of 13 autistic participants and half the NT group became aware they were controlling the lens, which may have implications in agency judgements.

#### INTERACT IN REALISTIC SCENARIOS

Simulations focused on skills requiring attention, such as vocational skills [89], [96], driving [70], [94], conversational skills [49], [75], [77], [97], [100] or interacting in public spaces.

Vocational skills (e.g., inspecting and sorting boxes in a warehouse) required attention to detail [89], [96]. Authors found that participants tended to ignore to assistive prompts and, instead, recommended highlighting objects.

Driving simulations highlighted salient stimuli (e.g., pedestrians, road signs) and used eye-tracking to track attention towards these [70], [94]. Participants showed significantly fewer driving errors [70], and better baseline performance was related to the rate of improvement in driving skills [94].

In other studies, participants wearing AR smartglasses that guided socially salient visual stimuli improved attention while interacting through conversation with an educator [77], [97], [100]. Another paper [49] found prompting participants with fading cues over faces was more effective than flying and exploding cues for directing attention; however only NT participants were involved.

During a public speaking task in which participants were required to attend to virtual peers while answering questions, [75] researchers examined moderating effects contributing to variances in attention and found social attention in autistic children were affected by IQ, social anxiety, and ADHD symptoms.

#### GAMIFICATION OF COGNITIVE TASKS

Many interventions involved gamifying repetitive tasks by having participants use motor movements to perform cognitive activities.

Participants used hand gestures to sort colored balls into corresponding boxes (e.g., [51], [80]) or pair matching items (e.g., words and definitions [99], objects [83]). Two other studies also required participants to move an object along a path using their hand [47] or a haptic arm [65].

Short-term improvements in sustained attention were observed during intervention sessions [51], [99] and long-term improvements in attention towards contextual information during a drag-and-drop task were maintained two weeks posttest [83]. Other long-term improvements in initiating and responding to joint attention cues using pointing, showing, and sharing skills during various scenarios were also made over a three-month period [86].

When comparing visuospatial attention between autistic and NT children, both groups demonstrated similar performance accuracy when raising their arms to hit balloons or bubbles. However, autistic children required longer time for successful attempts [95].

### *I. RQ9. WHAT WERE THE ATTENTION-RELATED OUTCOMES OF USING PHYSIOLOGICAL AND BEHAVIORAL MEASUREMENTS?*

A mini meta-analysis of physiological and behavioral metrics was conducted. Table 11 summarizes the nine papers that were included in this quantitative synthesis. Between-group studies with two groups of participants were included (autistic and age-matched group of neurotypical participants). The purpose of involving a NT group in these studies served to compare the implications of attention qualities (e.g., absolute or relative) on between-group differences of physiological or behavioral measures. Papers were omitted if they did not report statistical analyses or if the data was unrelated to attention qualities.

We reported the effect-sizes (partial eta squared) and F-values. Significance levels 0.01 (small) 0.06 (medium) and 0.14 (large) were taken from [110]. Relative attention demonstrated the lowest effect size (0.14), while absolute attention had an average effect size of 0.19. While reporting of results varied, we identified 15 p-values that were consistently reported. Furthermore, effect size ($n_p^2$) was calculated for single-way ANOVAs that reported F-values. p-values greater than 0.05 were plotted as 0.06, and those less than 0.05 were plotted as 0.04.

Absolute attention (i.e., total attention spent focusing on stimuli) of social stimuli and relative attention of social stimuli (i.e., the ratio of attention to stimuli compared to entire session) were identified as themes of qualities of attention from [103].

Five out of six of the absolute attention papers reported significant p-values ($p<0.05$) or trending towards significant levels (i.e., [82]). All relative attention literature failed to record strong evidence against null hypotheses [63], [67], [76], [103]. The remaining five articles recorded an insignificant p-value (i.e. [101]). However, this study also included the fewest participants.

**TABLE 11. Mini meta-analysis of studies featuring physiological measurement**

| Attention Qualities | Author (Year) | Physiological | Collection Method | *n* ASD Group | *n* NT Control Group | F-Value | T-Value | Z-Value | Effect Size | *p*-Value |
|---|---|---|---|---|---|---|---|---|---|---|
| Absolute Attention | [101] **Banire et al. (2019)** | Facial expression tracking | Frequency of expressions (social) | 4 | 4 | - | - | - | - | 0.4 |
| | [63] **Bekele et al. (2013)** | Eye-tracking | Number of fixations (social) | 10 | 10 | - | - | - | - | >0.05 |
| | [103] **Chen et al. (2019)** ***Exp. 1*** | Eye-tracking | Number of fixations (social)* | 10 | 13 | 5.69 | - | - | $n_p^2$=0.21 | <0.05 |
| | [82] **Grynszpan et al. (2012)** | Eye-tracking | Number of fixations (social) | 13 | 14 | 4.17 | - | - | $n_p^2$=0.14† | 0.0519 |
| | | Eye-tracking | Number of fixations* (non-social) | 13 | 14 | 6.03 | - | - | $n_p^2$=0.19† | 0.0213 |
| | | Eye-tracking | Sum of FD time* (social) | 13 | 14 | 7.28 | - | - | $n_p^2$=0.23† | 0.0018 |
| Relative Attention | [63] **Bekele et al. (2013)** | Eye-tracking | Proportion of FD (social) | 10 | 10 | - | - | - | - | >0.05 |
| | [67] **Bekele et al. (2014)** | Eye-tracking | Proportion of FD (social) | 10 | 10 | - | -1.47 | - | - | 0.1582 |
| | [103] **Chen et al. (2019)** ***Exp. 1*** | Eye-tracking | Proportion of FD | 10 | 13 | 3.27 | - | - | $n_p^2$=0.14 | 0.09 |
| | [76] **Feng et al. (2018)** | Eye-tracking | Proportion of FD time (social) | 26 | 26 | - | - | - | *d*=0.51 | > 0.05 |
| Other | [75] **Jarrold et al. (2013)** | Head-tracking | Frequency of head orientation* (social) | 37 | 54 | 6.71 | - | - | $n_p^2$=0.07† | <0.012 |
| | [78] **Jyoti et al. (2020)** | Pulse plethysmogram | Pulse rate* (social) | 20 | 20 | - | - | -0.41 | r=0.64 | <0.01 |
| | | Electrodermal activity | Tonic mean* (social) | 20 | 20 | - | - | -3.59 | r=0.57 | <0.01 |
| | [103] **Chen et al. (2019)** ***Exp. 1*** | Eye-tracking | Time to fixation* (social) | 10 | 13 | 13.39 | - | - | $n_p^2$=0.39 | <0.001 |
| | [62] **Pitskel et al. (2011)** | fMRI | Temporoparietal junction activity in response to gaze* (social) | 15 | 14 | 15.02 | - | - | $n_p^2$=0.36† | 0.000614 |

* Significant (p<0.05)
† Effect size (partial eta-squared) calculated from $n_p^2$=F**k*/(F**k*+df) for 95% CI

Overall, studies showed significant differences in attentiveness toward social ROIs (i.e., neutral virtual human face) between autistic and NT groups. On the other hand, high p-values for relative attention studies demonstrated a lack of significant differences in social viewing patterns between the two groups.

### J. RQ10. WHAT LIMITATIONS AND FUTURE RESEARCH AGENDAS WERE ADDRESSED IN THE PRIMARY LITERATURE?

In this section, we examine main limitations of included literature (Table 12). A primary challenge cited in the selected studies include small sample size, which ranged from fewer than 10 participants (e.g., n=1 [84], [92], [93], [97], n=3 [73], n=4 [40], [71], [101], n=5 [90], n=6 [65], n=8 [66], [100], n=9 [111] to over 10 participants (e.g., n=15 [71], n=17 [64]). Another commonly cited limitation was limited duration [40], [43], [71], [77], [86], [98], however, it was unclear what specific duration was perceived as limiting. For instance, one study included a single session lasting 10-20 minutes [77], while others involved two 180-minute sessions [43] or 14 five-minute sessions over five weeks [98].

TABLE 12. Main limitations of included studies.

| Limitations |
| --- |
| **1 Sample size**<br>• Gender parity<br>• Differences in participant profiles<br>• Selection biases during recruitment process |
| **2 Absence of comparison group** |
| **3 Equivalent metrics for calculating comparable performance across conditions** |
| **4 Problems with technology**<br>• Discomfort<br>• Development<br>• Technical issues |
| **5 Other**<br>• Practice effects |

Further limitations of small sample size include the lack of gender parity in the sample [85], [95] and differences in participant profiles [40]. Limitations in sample size may be partly due to selection biases during the recruitment process (e.g., only participants without intellectual disability [87]). Several studies also pointed out constraints to generalize results to the autistic population due to idiosyncratic differences between individuals with ASD [80], [88]. Task design appropriateness / ecological validity [63], [80]. Constraints associated with short duration involved inadequate exposure to technology or limited task repetition [96].

Another cited concern was the absence of a comparison group [94], [100] (e.g., no NT group [91], differences in gender ratio of the control group [82]) leading to an inability to examine differentiation effects. An additional recurrent limitation included the lack of equivalent metrics for calculating comparable performance across conditions [70], [95].

Other limitations mentioned include problems with technology (e.g., discomfort [94], development [81], technical issues [43]) and the possibility of practice effects [76], [95]. Several papers proposed conducting future research with extended intervention times [45], [77], [85], [88], [97], [100] to examine long-term effects of intervention [63] or transferability of skills [70] and incorporating physiological or behavioral tracking [48], [68], [93], [105].

## V. DISCUSSION

This section discusses research themes and sub-themes of key challenges (see Table 13) and makes recommendations on future research directions.

TABLE 13. Themes and sub-themes of literature

| Theme | Sub-theme |
| --- | --- |
| **A. Study Characteristics** | 1. Study Location |
| | 2. Participant Characteristics |
| **B. Autism Screening and Attention Measures** | 1. Assessment Tools |
| | 2. Physiological and Behavioral Metrics |
| **C. Considerations for Study Development** | 1. Task Appropriateness |
| | 2. Realistic Scenarios |
| | 3. Interactive Scenarios |
| | 4. Inattention and Distraction |
| | 5. Follow-up sessions |

### A. Study Characteristics

#### 1) STUDY LOCATION

Similar to previous reviews [112], the earliest study in this systematic review was published in 1996. The papers in this review originate from many of the same countries that consistently produce high outputs of autism research, including those in the U.S., Japan, China, and the U.K. [15]. This is also the first review of autism research using XR with a large number of studies derived from India, where autism diagnoses vary significantly across regions. There were no papers published from African countries, where autism remains largely underexplored [15]. It is crucial for future work to expand to different cultural communities and socio-economic groups as it would allow researchers to compare intervention strategies.

#### 2) PARTICIPANT CHARACTERISTICS

The majority of studies in the systematic review recruited participants between 3 and 18 years, although the prevalence of autistic adults is similar, if not greater [113]. Furthermore, adults may have learned to camouflage autistic characteristics, leading to underdiagnosis [114]. Male participants in these studies also greatly outnumbered female participants. These results are consistent to those attained by [33], who found many studies that included only autistic boys. The rate of males participating in studies greatly outweighed females and other genders. While this is in line with current gender ratios [15], females are often underdiagnosed due to the different ways ASD presents itself in this gender [115], [116]. For instance, autistic females are stronger in socio-cognitive skills [115], [117] and focus skills [118], which may contribute to differences in performance on attention tasks. It is, therefore, important to consider the effect of gender identity on user experience and efficacy as effective intervention methods from the primary literature may not transfer in the same way to female and non-binary populations.

An estimated 70% of autistic people have a comorbid psychiatric disorder, with 40% having two or more [1]. Approximately 25.7% of autistic people are also diagnosed with ADHD, and diagnosed anxiety disorders related to attention (e.g., anxiety disorder, obsessive-compulsive disorder) are also commonly linked with ASD (17.8%) [119]. This resulted in more generalizable and insightful outcomes. For instance, research that accommodated the effects of ADHD [75], [79], [100] observed impacts in autistic participants that were not present in groups with both ASD and ADHD. In contrast, some papers attempted to generalize results without considering the effects of learning disability or language disorder that resulted in inconclusive findings [85]. Given the substantial heterogeneity within ASD [78], [79], it is essential to examine the interplay of identities and comorbidities to better interpret and apply findings to wider intervention adoption and effects.

### *B. Autism Screening and Attention Measures*

#### 1) ASSESSMENT TOOLS

Several studies relied on standardized assessments with high reliability to evaluate outcomes. However, some of these failed to yield significant differences between pre- and post-test scores [85], produced inconsistent results [68], or resulted in ceiling effects [83]. Many studies also relied solely on self-developed assessments or rating scales [40] which raises questions on the validity of the results. Amaral *et al.* [72] developed a JA assessment, which did not yield any effect, but the standardized measures which were used as secondary outcome measures all yielded significant results. Researchers are encouraged to adopt appropriate, standardized quantitative measures consistently across XR studies to enable valid comparisons and more robust evaluations of attention and efficacy over time.

#### 2) PHYSIOLOGICAL AND BEHAVIORAL METRICS

Eye-tracking was the most widely used physiological measurement tool. Eye-gaze has consistently served as an accurate attention-tracking tool. Measurements, such as fixation duration, blink rate, and pupil dilation, are considered reliable indicators of psychological aspects including attention and engagement [120], [121]. This systematic review found studies that assessed the participant's fixation patterns and durations focused on social entities as regions of interest (e.g., facial features) [63], [92], [103]. Studies also tracked participant ability to follow the gaze of an avatar [45], [78]. Although not widely used across our featured studies, blink rate was found to be a more sensitive physiological index to indicate engagement with adaptive VR systems in comparison to pupil diameter [66]. Eye-tracking measurements and metrics can also differentiate between autistic and NT participants or other comorbid effects. For instance, two studies found significant group differences in attention towards human faces (i.e., number of fixations ($F(1, 25) = 4.17$, $p = 0.0519$ [82], $F(1, 21) = 5.69$, $p < 0.05$, $\eta_p^2 = 0.06$ [103]), fixation duration ($F(2, 47) = 6.94$, $p = 0.0023$) [82], and proportion of fixation duration $F(1, 21) = 5.69$, $p < 0.05$, $\eta_p^2 = 0.06$ [103]) between autistic and NT participants towards social stimuli. Furthermore, Mundy *et al.* [79] found autistic participants were less likely than those who were NT or with ADHD to exhibit large joint attention effects on information processing (Wilks' Lambda = 0.84, $F(2, 83) = 7.74$, $p < 0.001$, $\eta^2 = 0.16$).

Measurements of the autonomic nervous system can reflect affective responses not outwardly expressed [122], [123]. Jyoti and Lahiri [78] used these physiological indices to investigate participants' ease of understanding during joint attention tasks and found autistic participants showed higher variations in pulse rate compared to NT participants. Bekele *et al.* [63] provides supporting evidence that found such measurements (e.g., respiration, pulse, skin temperature) to be over 90% accurate in assessing stress during an attention task.

Head-position tracking can also be used to examine comorbid effects of ADHD. Jarrold *et al.* [75] found selective attention was lower in autistic participants with high anxiety and ADHD than those with low anxiety and ADHD based on how they oriented their heads during a public speaking task. However, other studies found it difficult to gauge whether participants were looking at ROIs correctly when turning their heads towards target stimuli [105].

fMRI studies with VR were primarily used to provide insight on differences in brain activity during social situations that may not otherwise be feasible to conduct in real-world settings. Two studies found different neural regions that showed levels of activation dissimilar to NT groups [62], [64]. For instance, direct gaze with an avatar elicited less activity in the region associated with social attention tasks related to the judgement of mental states (right temporoparietal junction) and more activity in the region related to selective attention (dorsolateral prefrontal cortex) [62], suggesting habitually reduced attention to socially salient stimuli.

Based on findings, physiological measurements provide the most salient insights. Eye-tracking, which appears to be the most used, is also the most accurate at measuring performance in attention tasks. fMRI, and ANS measurements (i.e., EEG, electrocardiogram, electrodermal activity, electromyogram, galvanic skin response, pulse plethysmogram, skin temperature, respiration) also provide an additional layer of understanding, although the ANS measurements may be the more feasible and cost-effective option of the three.

### *C. Considerations for Study Development*

#### 1) TASK APPROPRIATENESS

Complex tasks are challenging to assess and require reliable measurements [88]. Inconsistencies in the validity of attention tasks in pre- and post-test tasks further add to the difficulty of evaluating improvements on performance from interventions [80]. Thorough attention interventions with virtual scenarios highly relevant to the real environment produce transferable results [11].

#### 2) REALISTIC SCENARIOS

Lorenzo *et al.* [41] highlighted the importance of making VR scenarios as real as possible as this reduces adaptation time and increases skills learned during virtual role-playing that can then be transferred into the real-world. Cheng and Huang [86] also found children were more motivated to learn when they perceived the content to be closely related to their daily school activities.

Jyoti and Lahiri [78] modelled avatars after local Indian populations to facilitate the sense of real-life experience in JA tasks. One study found that when autistic children were able to customize an avatar to look like a real person they were familiar with, they were more likely to focus on context-relevant information [45]. This corresponds to another study in this review, which found that, in contrast to NT children, autistic children were unaffected by the uncanny valley effect. Interestingly, Kumazaki *et al.* [74] found autistic children responded more to virtual avatars and avoided looking at realistic humanoid robots, which suggests they could be affected by the uncanny valley effect for physical agents. While autistic users of VR report continuous distinction between real and virtual worlds, this aspect allows them to consider VEs to be safe places for training and learning [124]. This suggests a more significant potential for knowledge acquisition from realistic virtual tasks that can be applied to the real-world. In addition to customizable avatars, there are also research opportunities to investigate which additional aspects of VEs can be customized to facilitate a higher likelihood of transferable skills.

#### 3) INTERACTIVE SCENARIOS

Several articles provide promising results for gaze-contingent interfaces that integrate virtual tasks with eye-tracking metrics to facilitate engagement [66], [70]–[72]. Lahiri *et al.* [66] found that autistic participants were more likely to fixate on an avatar's face when using a gaze-contingent interface compared to a performance-based interface. Specifically, blink-rate was more receptive to engagement with the system. Additionally, autistic participants in [82] had similar results when presented with a gaze-contingent interface and shared a nearly equal number of fixations on an avatar's face as NT participants. There were no studies that assessed if progress from virtual interventions was translated into real-world settings, resulting in an opportunity to expand research in this area to explore transferability to real-life scenarios.

#### 4) INATTENTION AND DISTRACTION

Studies also focused on different ways to examine inattention using distractions. With the exception of two papers ([66], [74]), studies used distractions closely resembling those in real-life. When presented with non-social distractions, two studies found that autistic participants were less resistant to distractions than the NT control group based on tracked eye- and head-movements [65], [69]. However, some autistic participants were still able to perform similarly to NT participants [65]. In contrast, others performed significantly more poorly, despite having achieved similar scores to NT groups with traditional modalities (e.g., pen-and-paper) [69]. Interestingly, in vocational scenarios, participants in [89] reported feeling unaffected by distractors and found visually distracting environments to be more enjoyable [96]. However, these self-reports contradicted with performance scores which were lower than conditions with no distractions.

In contrast, autistic participants showed significant improvements in eye-tracking measures of inattention and performance following interventions in which interfaces highlighted relevant features in distracting driving scenarios [70], [94] or social contexts [77], [100]. Participatory design sessions also resulted in the development of an interface that filters irrelevant stimuli to augment attention towards important information [52]. These strategies of using visual supports are commonly used in autistic interventions, as they provide environmental structure and help users function more independently [125].

Tasks presenting users with rich scenarios and visual supports engage and train users to focus on essential information. Tracking eye- and head-movements provide valid measurements of improvements of inattention. Given that many inattention tasks are based on realistic situations, research efforts should assess how reactions to real-world distractions are affected by the intervention.

#### 5) FOLLOW-UP SESSIONS

Similar to Khowaja *et al.*'s [31] review, this systematic review found very few studies that involved maintenance phases or assessed transferability. Four studies that included maintenance sessions demonstrated how well participants retained the significant improvements they achieved during the intervention [71], [86]–[88]. However, the authors did not explore whether these improvements transferred to the participants' real-life. Only one study that conducted maintenance sessions outside the experimental setting confirmed that performance could be generalized to the participant's home [73]. Determining whether generalization is maintained during follow-up is considered a fundamental aspect of intervention studies in ASD [126] and should be factored into the study design.

## VI. CONCLUSION AND FUTURE WORK

This systematic review relied on six databases to search for peer-reviewed literature and may have missed relevant papers published in other databases (e.g., CINAHL, JSTOR, Web of Science). Furthermore, while the review procedure also attempted to encompass a range of papers, the search may have missed papers under other terms.

Authors focused solely on literature analysis and a basic meta-analysis to examine the corpus of papers. The current review did not include a quality assessment of individual studies or a formal meta-analysis. Future reviews would benefit from a risk of bias assessment and meta-analyses on evaluative research involving better-reported results to provide a more comprehensive understanding of research in this area.

The findings from this review provide substantial evidence for the effectiveness of using XR in attention-based interventions for autism to support autistic traits. However, the conclusions that can be drawn from this review are limited by inconsistent methodologies (e.g., evaluation measures) and inconclusive findings due to limitations such as small sample size from relevant studies to date. Despite these drawbacks, we were able to identify three principal gaps in the literature that could provide promising research directions. First, the lack of diversity in participants, namely underrepresented autistic populations such as adults, females and other genders, and individuals from lower-income countries could be solved by considering the effects of demographic differences to increase the generalizability of future findings to larger populations. Second, in many papers, there is a trade-off between using an assortment of standardized tasks and assessments and developing specific ones with untested reliability. However, we found benefits and insights were gained when appropriate assessments were combined with physiological measurements. Third, while many papers reported improvements in performance, future XR studies would benefit from further research to establish the specific aspects of interventions that are most effective in generating transferable skills. Furthermore, efforts to understand how performance in XR interventions is transferred to real-world scenarios is essential for the development of cost-effective and scalable applications. Lastly, understanding the varied interaction patterns of neurodivergent end users provides valuable insight to inform XR developers in the design of more inclusive applications and, in doing so, would enhance user experience for a more diverse range of users.

XR offers an accessible and engaging way to support attention differences in autism. With 1.2% of the global population diagnosed as autistic, and potentially even more undiagnosed, it is essential for research to consider how to accommodate autistic trait differences to create accessible and effective therapy solutions and supportive interfaces for engaging comfortably in learning, playing, and interacting.

## ACKNOWLEDGEMENT

This work was supported in part by the Foreign, Commonwealth & Development Office through the AT2030 Programme (www.AT2030.org).

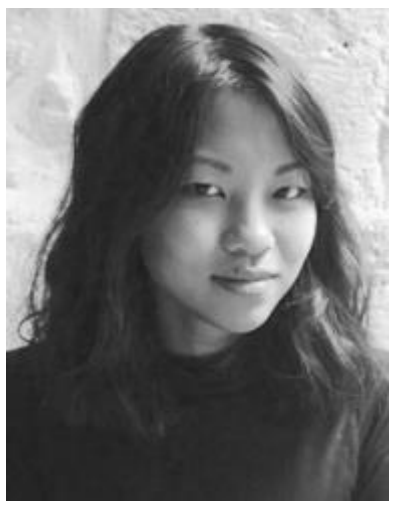

**KATHERINE WANG** received a B.Sc. degree in Psychology from the Pennsylvania State University in 2018 and a M.Sc. degree in Disability, Design and Innovation from University College London (UCL), London, U.K. in 2020. She is currently pursuing a Ph.D. degree in Human-Computer Interaction at the Global Disability Innovation Hub and UCL Interaction Centre (UCLIC). Her research interests include using virtual reality as an assistive technology and understanding how biofeedback signals can be incorporated into the design of virtual environments to support social connections.

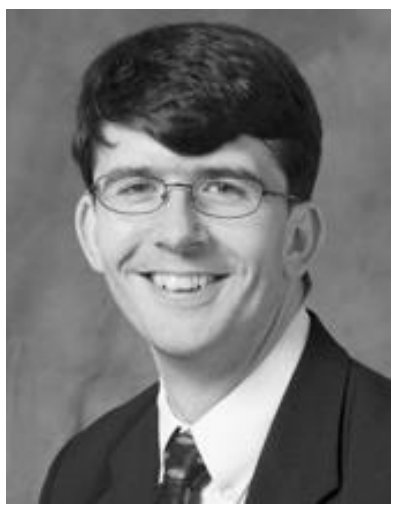

**PROF. SIMON J. JULIER** is currently a Professor with the Vision, Imaging and Virtual Environments Group, Department of Computer Science, University College London (UCL), London, U.K.

Before joining UCL, he worked for nine years with the 3D Mixed and Virtual Environments Laboratory, Naval Research Laboratory, Washington, DC, U.S.A. He has worked on a number of projects, including the development of systems for sports training, coordinated search, and rescue with swarms of UAVs, remote collaboration systems, enhanced security management systems for refugee camps, and sea border surveillance in the presence of small targets. His research interests include distributed data fusion, multitarget tracking, nonlinear estimation, object recognition, and simultaneous localization and mapping.

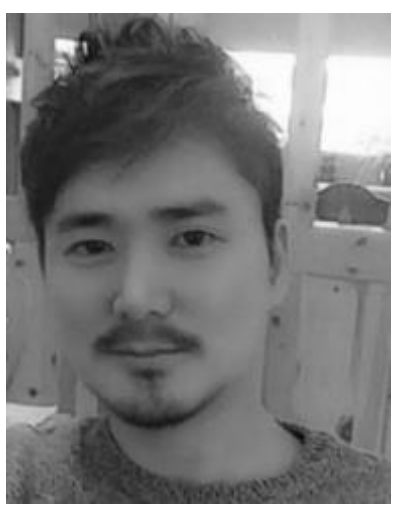

**DR. YOUNGJUN CHO** received a Ph.D. degree in computational physiology from Faculty of Brain Sciences, University College London (UCL), London, U.K.

He is currently an Associate Professor with the Global Disability Innovation Academic Research Centre (GDI-ARC) and UCL Interaction Centre (UCLIC), Department of Computer Science, UCL. He explores, builds, and evaluates novel technologies for the next generation of AI-powered physiological computing* that helps boost disability technology innovation. *His definition of physiological computing is technology that helps us listen to our bodily functions, psychological needs and adapts its functionality.